\documentclass[aps,prd,twocolumn,showpacs]{revtex4}

\usepackage{amsmath}
\usepackage{amsfonts}
\usepackage{subeqnarray}
\usepackage{graphicx}	
\usepackage{epstopdf}

\begin{document}


\title{Non-Abelian Chern-Simons-Higgs vortices with a quartic potential}

\author{J. L. Bl\'azquez-Salcedo$^{1}$, L. M. Gonz\'alez-Romero$^{1}$, F. Navarro-L\'erida$^{2}$, and D. H. Tchrakian$^{3,4}$}

\affiliation{{$^{1}$ Departamento de F\'{\i}sica Te\'orica II, Universidad Complutense de Madrid, E-28040 Madrid, Spain}}
\affiliation{{$^{2}$ Departamento de F\'isica At\'omica, Molecular y Nuclear,  Universidad Complutense de Madrid, E-28040 Madrid, Spain}}
\affiliation{{$^{3}$ School of Theoretical Physics, Dublin Institute for Advanced Studies, 10 Burlington Road, Dublin 4, Ireland }}
\affiliation{{$^{4}$ Department of Computer Science, NUI Maynooth, Maynooth, Ireland}}

\date{\today}

\begin{abstract}
We have constructed numerically non-Abelian vortices in an $SU(2)$
Chern-Simons-Higgs theory with a quartic Higgs potential. We have analyzed
these solutions in detail by means of improved numerical codes and found some unexpected features we did not find when a sixth-order Higgs potential was used. The generic
non-Abelian solutions have been generated by using their corresponding Abelian
counterparts as initial guess. Typically, the energy of the non-Abelian
solutions is lower than that of the corresponding Abelian one (except in
certain regions of the parameter space). Regarding the angular momentum, the
Abelian solutions possess the maximal value, although there exist non-Abelian
solutions which reach that maximal value too. In order to classify the
solutions it is useful to consider the non-Abelian solutions with
asymptotically vanishing $A_t$ component of the gauge potential, which may be labelled by an integer number $m$. For vortex number $n=3$ and above, we have found uniqueness violation: two different non-Abelian solutions with all the global charges equal. Finally, we have investigated the limit of infinity Higgs self-coupling parameter and found a piecewise Regge-like relation between the energy and the angular momentum.

\end{abstract}


\maketitle

\section{Introduction}
Vortices on $\Re^2$ have attracted interest since a very long time. They arise in spontaneously broken gauge theories in two dimensions and possess a quantized magnetic flux due to their topological properties. When a Chern-Simons (CS) term is added to the action, vortices acquire electric charge while keeping a finite energy \cite{Vega_Schaposnik_1}. The inclusion of CS terms in Higgs models in 2+1 dimensions was motivated by the discovery in \cite{Deser}  of topologically massive non-Abelian Yang-Mills (YM) theories augmented by a CS term, where the CS provides a gauge invariant mechanism of mass generation.

Vortex solutions in an Abelian Chern-Simons-Higgs (CSH) theory were studied in \cite{Hong,JW} by Hong, Kim, and Pac and Jackiw and Weinberg (HKP-JW), independently. There, a sixth-order Higgs potential was used to ensure self-duality. Non-Abelian generalizations of these solutions were considered later \cite {Vega_Schaposnik_1,Vega_Schaposnik_2,Kumar_Khare} with a simple gauge group ($SU(2)$ and $SU(N)$). In contrast to our model, these models feature at least two adjoint representation Higgs fields in addition to other scalar multiplets. Owing to that these solutions are topologically stable.

The non-Abelian generalization of the Abelian vortices introduced in  \cite{Hong,JW} was presented in \cite{CSH}, where only one Higgs field is considered. There a sixth-order Higgs potential is employed. In this paper we investigate an $SU(2)$ CSH model with the Higgs field in the adjoint representation but using the standard quartic potential for the Higgs field. Due to that we will not have self-dual solutions in the Abelian sector of the theory. By using improved numerics we are able to analyze the solutions very accurately and explore the limit of large Higgs self-coupling constant. 

The paper is organized as follows: in next section we present the model. Then in Sec. III we introduce the Ansatz and the gauge choice we will employ and derive the field equations for that Ansatz in Sec. IV. Due to their special relevance in the construction of the vortex solutions we devote Sec. V to the Abelian case. We carry out the numerical construction of the non-Abelian vortex solutions in Sec. VI and summarize our results in Sec. VII.

\section{Field equations}
We will use the Lagrangian density 
\begin{eqnarray}
\mathcal{L}&=&\frac{\kappa}{2}\varepsilon^{\rho \mu \nu}
Tr\left[A_{\rho}(F_{\mu\nu}-\frac{2}{3}A_{\mu}A_{\nu})\right] \nonumber\\ 
&+& Tr[D_{\mu}\Phi D^{\mu}\Phi] -
V(\Phi),
\label{lagrangian}
\end{eqnarray}
where $\varepsilon^{\alpha \mu \nu}$ is the three dimensional Levi-Civita
tensor, $A_{\mu}$ is an $SU(2)$ gauge potential, and $\Phi$ is the Higgs field in the
adjoint representation. We have defined the gauge field by
\begin{equation}
F_{\mu \nu}=\partial_{\mu}A_{\nu} - \partial_{\nu}A_{\mu} + [A_{\mu},A_{\nu}],
\end{equation}
and the gauge covariant derivative by
\begin{equation}
D_{\mu}=\partial_{\mu} + [A_{\mu},\circ].
\end{equation}
Both the gauge potential and the Higgs field can be written in terms of combinations of $su(2)$ matrices
\begin{eqnarray}
&&A_{\mu}=A_{\mu}^{a}T_{a},\\
&&\Phi=\Phi^{a}T_{a},
\end{eqnarray}
with $T_{a}=\frac{1}{2i}\tau_{a}$ (a=1,2,3), $\{\tau_{a}\}$ being the Pauli matrices.

The Lagrangian density Eq.~(\ref{lagrangian}) describes the coupling between the gauge field
and the Higgs field. There is no dynamical term for the gauge
field, we are considering a CS coupling term though. For the potential $V(\Phi)$ we will employ the standard quartic symmetry-breaking Higgs self-interaction potential
\begin{equation}
V(\Phi)=-(4\lambda)^{2}Tr\left[\frac{1}{4}v^{2} + \Phi^{2}\right]^{2} .
\end{equation}

By taking variations of the Lagrangian with respect to the gauge potential
and the Higgs field, we obtain the Euler-Lagrange equations of motion. These
general equations read 
\begin{eqnarray}
&&D_{\mu}(D^{\mu}\Phi)-2(4\lambda)^{2}\Phi(\frac{1}{4}v^{2} + \Phi^{2}) =0
\label{eqn_Phi_gen},  \\
&&\frac{\kappa}{2}\varepsilon^{\alpha \mu \nu} F_{\mu \nu} + [\Phi,D^{\alpha}\Phi]=0. \label{eqn_F_gen}
\end{eqnarray} 
Note that Eq.~(\ref{eqn_Phi_gen}) gives the Higgs field
dynamics and Eq.~(\ref{eqn_F_gen}) can be seen as a set of constraint
equations that bounds the gauge field to the Higgs field.

From the Lagrangian Eq.~(\ref{lagrangian}) we may also obtain the stress-energy tensor:
\begin{eqnarray} 
T_{\mu \nu} &=&
\eta_{\mu \nu}Tr\left[(D_{\alpha}\Phi)^{2}\right] - 2Tr[D_{\mu}\Phi D_{\nu}\Phi] \nonumber \\ &+& 16\lambda^{2}\eta_{\mu
  \nu}Tr\left[\frac{1}{4}v^{2} + \Phi^2\right],
\end{eqnarray}
from which we will obtain the energy and the angular momentum of the vortex configurations. Here $\eta_{\mu \nu}$ denotes the Minkowski metric.

\section{Ansatz and gauge choice}
We will restrict to a rotationally symmetric Ansatz for the Higgs field and the Yang-Mills (YM) connection. They may be written in the following form \cite{CSH}
\begin{widetext}
\begin{eqnarray}
&&\Phi    =  \phi^{(3)} T_{r}^{(n)} + \phi^{(4)} T_{\varphi}^{(n)} - \phi^{(5)}
T_{3}, \label{ansatz_phi}\\
&& A_{t}  =  \chi^{(3)} T_{r}^{(n)} + \chi^{(4)} T_{\varphi}^{(n)} -   \chi^{(5)}
 T_{3}, \label{ansatz_At}\\
&& A_{1}  = - \left[\frac{\xi^{(3)}}{r}\hat{x}_{2} + A^{(3)}_{r} \hat{x}_{1}\right] T_{r}^{(n)} -
 \left[\frac{\xi^{(4)}}{r}\hat{x}_{2} + A^{(4)}_{r} \hat{x}_{1}\right] T_{\varphi}^{(n)}  +
 \left[A^{(5)}_{r} \hat{x}_{1} + \left(\frac{\xi^{(5)} + n}{r}\right) \hat{x}_{1}\right]
 T_{3}, \label{ansatz_Ax}\\
&& A_{2}  = \left[\frac{\xi^{(3)}}{r}\hat{x}_{1} - A^{(3)}_{r} \hat{x}_{2}\right] T_{r}^{(n)} +
 \left[\frac{\xi^{(4)}}{r}\hat{x}_{1} - A^{(4)}_{r} \hat{x}_{2}\right] T_{\varphi}^{(n)}  +
 \left[A^{(5)}_{r} \hat{x}_{2} - \left(\frac{\xi^{(5)} + n}{r}\right)
   \hat{x}_{2}\right] T_{3} \label{ansatz_Ay},
\end{eqnarray}
\end{widetext}
where we have defined $n_{1}=\cos n\varphi$, $n_{2}=\sin n\varphi$,
$\hat{x}_{1}=\cos \varphi $, and $\hat{x}_{2}=\sin \varphi$, and the $su(2)$-valued matrices
\begin{eqnarray}
&&T_{r}^{(n)}=\cos n\varphi \ T_{1} + \sin n\varphi \ T_{2} , \\
&&T_{\varphi}^{(n)}=\cos n\varphi \ T_{2} - \sin n\varphi \ T_{1}.
\end{eqnarray}
$n$ denotes an integer number, representing the winding (vortex) number.
The Ansatz functions $\phi^{(i)}, \chi^{(i)},  \xi^{(i)}$, and $A^{(i)}_{r}$
$(i=3,4,5)$ depend on the radial coordinate $r$ only . 

Equations (\ref{ansatz_phi}-\ref{ansatz_Ay}) represent the most
general expression for the Ansatz, which describes configurations of $n$ vortices pinned up at the origin $(r=0)$. But we still have a complete $SU(2)$ gauge symmetry that can be used to simplify the Ansatz.

Part of that gauge freedom may be removed by setting  $A_{r}^{(i)}(r)=0$ . Furthermore, we simplify a bit more the Ansatz by introducing the consistent truncation $\phi^{(4)}=0$, $\chi^{(4)}=0$, and $\xi^{(4)}=0$ \cite{CSH}.

Then, the only functions we have in our truncated Ansatz are: $\phi^{(3)}=vh,
\phi^{(5)}=vg,\xi^{(3)}=c, \xi^{(5)}=a,
\chi^{(3)}=\frac{v^{2}}{\kappa}d,\chi^{(5)}=\frac{v^{2}}{\kappa}b$, where $a$,
$b$, $c$, $d$, $g$, and $h$ are functions of $r$. In these variables the Ansatz reads
\begin{eqnarray}
\Phi   &=& vh T_{r}^{(n)} - vg T_{3}, \label{ansatz_final_phi}\\
 A_{t} &=& \frac{v^{2}}{\kappa}d T_{r}^{(n)} - \frac{v^{2}}{\kappa}b T_{3}, \label{ansatz_final_At}\\
 A_{r} &=& 0, \label{ansatz_final_Ar}\\
 A_{\varphi} &=& -\frac{2c}{r}\cos \varphi  \sin \varphi  T_{r}^{(n)}
 \nonumber  \\ &-&
 \frac{2(a+n)}{r}\cos \varphi  \sin \varphi 
 T_{3},  \label{ansatz_final_Avarphi}
\end{eqnarray}
where we have introduced the $(r,\varphi)$ components of the gauge connection
\begin{eqnarray}
&&A_{r}=\cos \varphi A_{1} + \sin \varphi 
A_{2} , \\
&&A_{\varphi}=\cos \varphi A_{2} - \sin \varphi A_{1}.
\end{eqnarray}

\section{Differential Equations, Energy, and Angular Momentum of the Solutions} 
If we redefine the parameters of the theory the following way
\begin{equation}
\kappa=\frac{v^{2}}{\beta}, \ \ \ \lambda=\frac{\beta\mu}{2v},
\end{equation}
and rescale the radial coordinate by $r\rightarrow
\beta r$,  the only parameter of the theory that can be found
explicitly in the differential equations  is $\mu$
(the scaled Higgs self-coupling parameter). In these rescaled variables the field equations read
\begin{eqnarray}
a_{,r}&=&(gd-hb)hr \label{ed_a},\\
b_{,r}&=&-(ah-gc)h/r \label{ed_b},\\
c_{,r}&=&-(gd-hb)gr \label{ed_c},\\
d_{,r}&=&(ah-gc)g/r \label{ed_d},\\
h_{,rr}&=&-\frac{1}{r}h_{,r} + \frac{a}{r^{2}}(ah-gc) \nonumber
 \\ &+& b(gd-bh) + 2\mu^{2}h(g^{2}+h^{2}-1) \label{ed_h},\\
g_{,rr}&=&-\frac{1}{r}g_{,r} - \frac{c}{r^{2}}(ah-gc) \nonumber
 \\ &-& d(gd-bh) + 2\mu^{2}g(g^{2}+h^{2}-1) \label{ed_g},
\end{eqnarray}
together with the constraint  
\begin{equation}
rh g_{,r}-rg h_{,r}+ad-bc = 0.
\label{ed_constrain}
\end{equation}
Here the subindex ${}_{,r}$ denotes the derivative with respect to the
rescaled radial coordinate $r$.

The constraint equation Eq.~(\ref{ed_constrain}) is compatible with the system of differential equations
Eqs.~(\ref{ed_a}-\ref{ed_g}), so we can take Eqs.~(\ref{ed_a}-\ref{ed_h})
and Eq.~(\ref{ed_constrain}) as the minimal system of equations of the problem.


The total energy of a given solution can be calculated from the $(t,t)$
component of the stress-energy tensor:
\begin{equation}
E=\int_{\Re^{2}}T_{t t}=2\pi\int_{0}^{\infty}dr rT_{t t} \label{energy_int}.
\end{equation}
Note that $2\pi rT_{t t}=H$, the Hamiltonian. For the Ansatz we are considering, the
expression of the Hamiltonian is:
\begin{eqnarray}
H&=&\pi r \left[ \left( g_{,r} \right)^{2} + \left( h_{,r}
  \right)^{2} \right] + \frac{\pi}{r}(gc-ha)^{2} \nonumber \\ &+& \pi r(bh - gd)^{2} +
\mu^{2}\pi r (1-g^{2}-h^{2})^{2}.
\end{eqnarray}
Using the expression of the total energy of the configurations, we may obtain
the set of boundary conditions that allows us to generate solutions. Since we
are interested in vortex solutions, we must impose regularity of the
solutions at the origin and a finite value of the energy.

The expansion at the origin of a general vortex solution with vorticity $n$
reads: 
\begin{eqnarray}
&&a = -n -\tfrac{h_{n}^{2}}{2(n+1)}\left( b_{0} + g_{0}^{2}\right)r^{2n+2} + O(r^{2n+4}), \label{exp_a_origin}\\
&&b = b_{0} + \frac{h_{n}^{2}}{2}r^{2n} + O(r^{2n+2}), \\
&&c = \frac{b_{0}+g_{0}^{2}}{n+2}g_{0}h_{n}r^{n+2} + O(r^{n+4}), \\
&&d = -g_{0}h_{n}r^{n} + O(r^{n+2}), \\
&&g = g_{0} + \mu^{2}\frac{g_{0}}{2}(g_{0}-1)(g_{0}+1) r^{2} + O(r^{4}), \\
&&h = h_{n}r^{n} + O(r^{n+2}), \label{exp_h_origin}  
\end{eqnarray}
and all the higher order terms can be written in terms of $g_{0}, b_{0}$, and
$h_{n}$. Note that, although the vorticity number does not appear explicitly in the
equations, it is a fundamental parameter in the expansion at the origin.
 
Finiteness of the energy Eq.~(\ref{energy_int}) imposes the following
asymptotic values for the functions:
\begin{eqnarray}
&&\lim_{r \to +\infty}a=p_{1}\cos \alpha \ , \ \ \lim_{r \to
  +\infty}b=p_{2}\cos \alpha \label{exp_a_infty},\\ 
&&\lim_{r \to +\infty}c=p_{1}\sin
\alpha \ ,\ \ \lim_{r
  \to +\infty}d=p_{2}\sin \alpha ,\\  
&&\lim_{r \to +\infty}g=\cos
\alpha \ ,\ \  \lim_{r
  \to +\infty}h=\sin \alpha .  \label{exp_h_infty}
\end{eqnarray} 
$p_{1}$ is related to the amplitude of the electric isotriplet $\vec\chi$, and $p_{2}$
to the amplitude of $\vec\xi$. $\alpha$ is related to the angle between the directions of
the non-Abelian isotriplets and their Abelian counterparts. 

From all the parameters involved in the expansion at the origin and the
behavior at infinity, $g_{0}$, $b_{0}$, $h_{n}$, $p_{1}$, $p_{2}$, and $\alpha$, it
can be numerically proven that only one is free and the remaining ones are
numerically fixed by the system. In our computations we have chosen either
$p_{1}$ or $p_{2}$ as the free numerical parameter, depending on the numerical
convenience of one or the other.


Another physical quantity that will turn out to be useful in the analysis of
the solutions is the angular momentum $J$. From the $(t,\varphi)$ component of
the stress-energy tensor we obtain the 
total angular momentum of the configuration:
\begin{equation}
J=\int_{\Re^{2}}T_{t \varphi}=2\pi\int_{0}^{\infty}dr\ rT_{t \varphi} ,
\end{equation}
%
which results to be
\begin{equation}
J=2\pi\int_{0}^{\infty}dr\ r(gd-hb)(gc-ha).
\end{equation}

Using Eqs.~(\ref{ed_a}) and (\ref{ed_c}) and the values of the
functions at the origin and infinity, the total angular momentum of a
configuration may be shown to depend only on the vorticity
$n$ and the asymptotic parameter $p_{1}$ \cite{CSH}
\begin{eqnarray}
J&=&-2\pi\int_{0}^{\infty}dr \  (c c_{,r} + a a_{,r}) 
= \pi(n^{2} - p_{1}^{2}).
\label{total_angular_momentum}
\end{eqnarray}

\section{Abelian case}

Due to their essential role in the construction of the non-Abelian vortices, we
will analyze the embedded Abelian solutions. The Ansatz,
Eqs.~(\ref{ansatz_final_phi}-\ref{ansatz_final_Avarphi}), becomes Abelian when 
\begin{equation}
c=d=g=0.
\end{equation}
%

The equations for the Abelian case are greatly simplified. It should be noticed
that the constraint
equation Eq.~(\ref{ed_constrain}) is identically satisfied in this Abelian case,
so the minimal system of equations reduces to
\begin{eqnarray}
a_{,r}&=&-bh^{2}r \label{ed_a_ab},\\
b_{,r}&=&-ah^{2}/r \label{ed_b_ab},\\
h_{,rr}&=&-\frac{1}{r}h_{,r} + \frac{a^{2}}{r^{2}}h
- b^{2}h + 2\mu^{2}h(h^{2}-1) \label{ed_h_ab}.
\end{eqnarray}
These equations are the analogue to the equations used by HKP-JW \cite{Hong,JW}, with a
quartic potential instead. The Hamiltonian for this Abelian configurations is
\begin{equation}
H=\pi r \left( h_{,r}
  \right)^{2} + \frac{\pi}{r}(ha)^{2} + \pi r(bh)^{2} +
\mu^{2}\pi r (1-h^{2})^{2}.
\label{hamiltonian_Abelian}
\end{equation}
Now imposing regularity at the origin and finite energy, we obtain that there is no free integration parameter for
the Abelian configurations, once $\mu$ is given. In fact, comparing the parameters with the general non-Abelian case, one finds that
\begin{equation}
p_{1}=p_{2}=0, \alpha=\pi/2 ,
\label{Abelian_parameters}
\end{equation}
for Abelian solutions. Note that due to Eq.~(\ref{total_angular_momentum}), the Abelian solutions
possess maximal angular momentum. 

These embedded Abelian solutions constitute the starting point in the
construction of non-Abelian vortices, since the non-Abelian $p_2 =0$ branches
bifurcate from the Abelian branch at certain values of the rescaled Higgs
self-coupling parameter $\mu$. The remaining  non-Abelian $p_2 \neq 0$
solutions may be computed from these non-Abelian $p_2=0$ counterparts.   

\section{Numerical Results}
The complexity of the system of equations Eqs.~(\ref{ed_a}-\ref{ed_h}) and
(\ref{ed_constrain}) prevents us from using analytical methods to solve it. On
the contrary, numerical schemes may be successfully employed. The set of boundary
conditions for numerics can be easily derived from the expansion at the origin
Eqs.~(\ref{exp_a_origin}-\ref{exp_h_origin}) and asymptotic behavior
Eqs.~(\ref{exp_a_infty}-\ref{exp_h_infty}) of the functions. In fact, several choices are
possible that ensure convergence of the codes.

Compared to our previous paper \cite{CSH}, we have improved our numerical accuracy, which has allowed us to analyze vast regions of the parameter space in detail, including the limit $\mu \to \infty$. We have applied a collocation method for boundary-value ordinary differential equations, equipped with an adaptive mesh selection procedure \cite{colsys}. Typical mesh sizes include $10^3$ - $10^4$ points. The solutions have a relative
accuracy of $10^{- 8}$.

After a detailed analysis of the equations, one finds that for a fixed integer
value of the vortex number $n$ and a nonvanishing real value of the Higgs
self-coupling constant $\mu$, the regular solutions to
Eqs.~(\ref{ed_a}-\ref{ed_h}) and (\ref{ed_constrain}) depend on just one
numerical parameter which we have chosen to be either $p_1$ or $p_2$. Usually
$p_2$ is the most efficient numerical parameter, but in certain regions of the
parameter space the system becomes extremely sensitive to changes in $p_2$ and
using $p_1$ as the free parameter improves the efficiency of the numerical
codes. 

Our procedure to generate non-Abelian vortices in the $\{n,\mu,p_{2}\}$
parameter space was as follows: for fixed integer $n$, we start from a small
value of $\mu$ and generate the corresponding Abelian solution ($p_2=0$); this
may be done easily by using a shooting method; after that, the $p_2$ parameter is
moved from zero while keeping $n$ and $\mu$; this generates non-Abelian
solutions, as $c$, $d$, and $g$ functions get excited; once the non-Abelian
solutions are generated, one may study the parameter space moving $p_2$ and
$\mu$. In order to generate solutions with a different value of the vortex
number $n$, one has to start from the corresponding Abelian solution as $n$
cannot be varied smoothly from an integer value to another.

The profiles of the functions $a$, $b$, $c$, $d$, $g$, and $h$ for a typical
non-Abelian solution ($n=1$, $\mu=50$, $p_2=0.6$) are shown in Fig.~1. The deviation of $c$, $d$, and $g$ from zero is clearly seen, which remarks
the non-Abelian nature of the solution.

\begin{figure}
\includegraphics[angle=-90,width=0.5\textwidth]{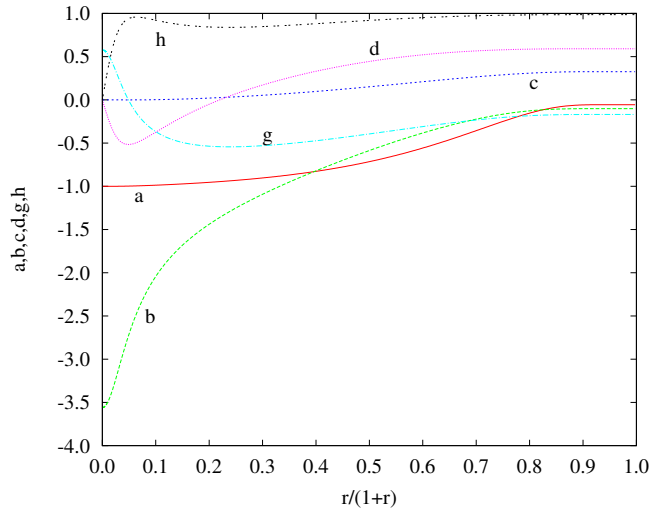}
\caption{Functions $a$, $b$, $c$, $d$, $g$, and $h$ for a typical non-Abelian
  solution ($n=1$, $\mu=50$, $p_2=0.6$).}
\label{functions_50_p2_0.6}
\end{figure}

As one moves $p_2$ away from zero, the solutions become more non-Abelian. This
non-Abelianess has consequences on the global charges such as the energy and
the angular momentum. In Fig.~2 we exhibit the energy $E$ and the angular
momentum $J$ versus the asymptotic
parameter $p_2$ for $n=1$, $\mu=50$ solutions. We observe that the
energy decreases as we separate from the Abelian solution. In fact, for wide
ranges of $\mu$ the corresponding Abelian solution always has the largest
 value of the energy along the branches with fixed $n$ and $\mu$,
although for small values of $\mu$ there exist non-Abelian solutions with
energy greater than that of their Abelian counterpart. For the angular
momentum the Abelian solutions always have the maximal value, as can be
easily seen from Eq.~(\ref{total_angular_momentum}) ($p_1=0$ for Abelian
solutions).

\begin{figure}
\includegraphics[angle=-90,width=0.5\textwidth]{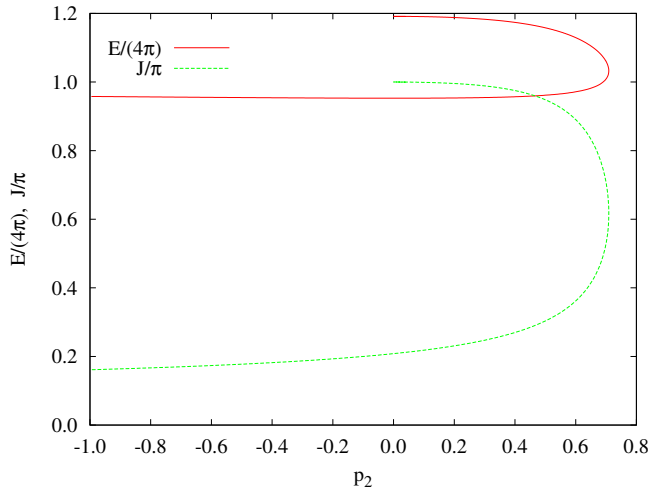}
\caption{Energy $E$ and angular momentum $J$ versus the $p_{2}$ parameter for $n=1$, $\mu=50$ vortices.}
\label{fig_p2_E_J_50}
\end{figure}

The theory possesses the symmetry $p_2  \to -p_2$, leaving the
global charges invariant. Then, the mirror image of the curves presented in Fig.~2 might
be plotted, although we will not include those mirror images in our figures
for the sake of clarity.

The non-Abelian branch may be extended by moving $p_2$ until the limit $|p_2|
\to 1$ is reached. In that limit the solutions tend pointwise to a trivial
solution \cite{CSH}.

\subsection{Abelian and non-Abelian $p_2=0$ branches}
\begin{figure}
\includegraphics[angle=-90,width=0.5\textwidth]{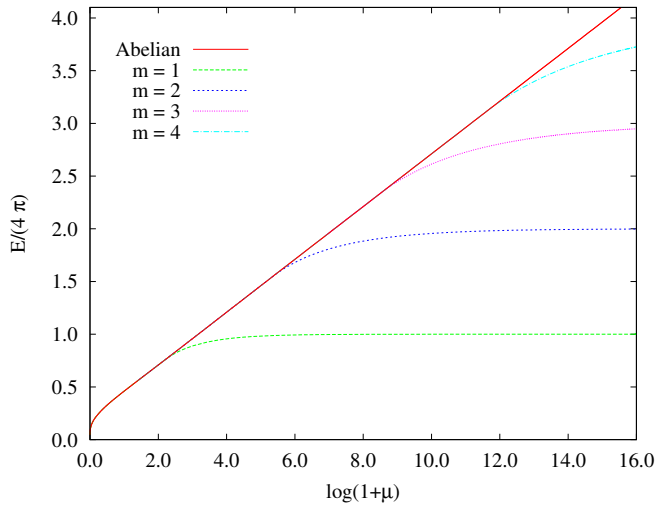}
\caption{Energy $E$ versus the Higgs potential coupling constant $\mu$ for CSH
  vortices with $n=1$, $p_{2}=0$.}
\label{fig_E_branches_n1}
\end{figure}
\begin{figure}
\includegraphics[angle=-90,width=0.5\textwidth]{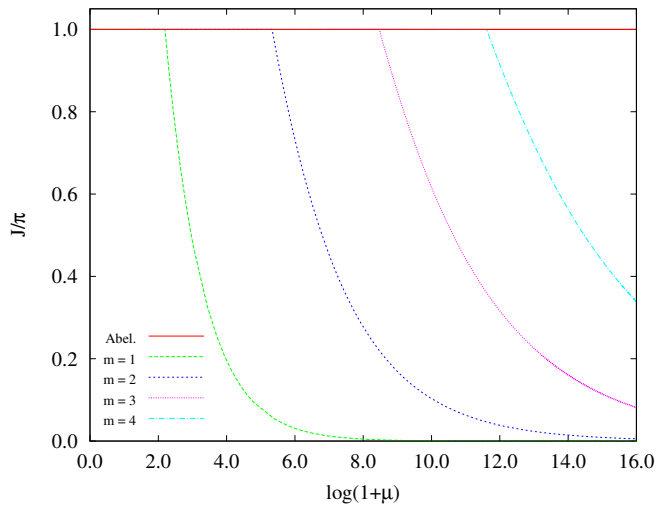}
\caption{Angular momentum $J$ versus the Higgs potential coupling constant $\mu$ for CSH
  vortices with $n=1$, $p_{2}=0$.}
\label{fig_J_branches_n1}
\end{figure}
\begin{figure}
\includegraphics[angle=-90,width=0.5\textwidth]{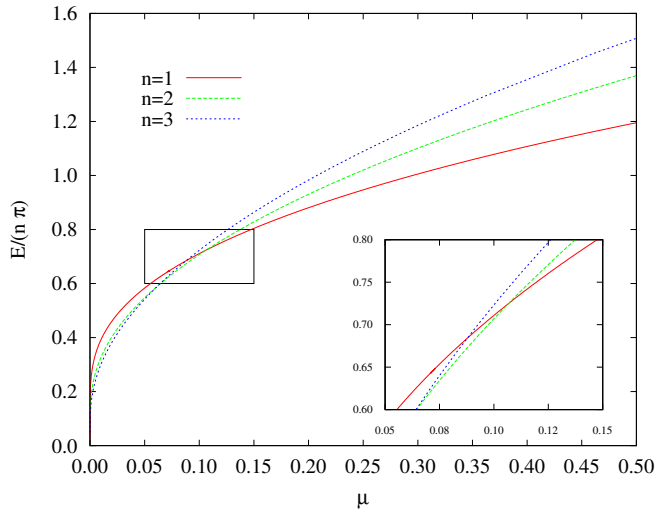}
\caption{Energy per vortex $E/n$ versus the Higgs potential coupling constant $\mu$ for CSH
  vortices with $n=1,2,3$, $p_{2}=0$.}
\label{fig_self_duality}
\end{figure}

Although Abelian vortices imply $p_2=0$, the opposite does not hold. In fact, in
Fig.~2 the existence of one non-Abelian solution with
$p_2=0$ is clearly seen. These non-Abelian $p_2=0$ solutions will be shown to be crucial in
understanding of the structure of the solution space.

For any nonvanishing integer value of $n$ and nonvanishing real value of $\mu$
there exists a unique Abelian vortex solution. For low values of $\mu$ this
solution is the only one with $p_2=0$. As $\mu$ is increased non-Abelian
$p_2=0$ solutions branch off the Abelian ones. These new non-Abelian $p_2=0$
branches can be enumerated by using an integer number $m$, that labels
them. For fixed $\mu$ the number of non-Abelian $p_2=0$ branches is finite,
this number increasing with increasing $\mu$ (in the limit $\mu \to \infty$
the number of these branches becomes infinite.) The values of $\mu$ where the
non-Abelian $p_2=0$ branches bifurcate from the Abelian one depend on the
vortex number $n$, and they are roughly equidistant on a logarithmic scale for
$\mu$. For example, for $n=1$, the first
non-Abelian $p_2=0$ branching points are $\mu=8.023$ $(m=1)$,
$\mu=2.063\cdot10^{2}$ $(m=2)$, $\mu=4.769\cdot10^{3}$ $(m=3)$, and
$\mu=1.108\cdot10^{5}$ $(m=4)$.

The general structure of these $p_2=0$ solutions is exhibited in Fig.~3 for
$n=1$ solutions. As happened for their YMH analogues \cite{YMH} , the energy of
non-Abelian $p_2=0$ solutions is always smaller than that of the corresponding Abelian
solution. Notice this was not the case for CSH solutions with a sixth-order potential
\cite{CSH}. Then, this fact seems to be a consequence of the quartic Higgs
potential.  For other values of the vorticity number, the
behaviors of the Abelian and non-Abelian branches are quite similar.

In the limit $\mu \to \infty$ the energy of the non-Abelian $p_2=0$ solutions
tends to $E_{n, m}\to 4\pi mn$ asymptotically. The energy of the Abelian configuration diverges logarithmically, though \cite{Burzlaff}.

The angular momentum $J$ of these non-Abelian $p_2=0$ configurations is also below the
angular momentum of their Abelian counterparts, and tends asymptotically to
zero in the limit $\mu \to \infty$. This is exhibited in
Fig.~\ref{fig_J_branches_n1}, where $J$ is plotted versus $\mu$ both for
Abelian and for non-Abelian $p_2=0$ solutions.

The lack of self-duality of the Abelian configurations of this theory can be
shown by representing the energy per vortex for some Abelian configurations
for different vortex numbers. In Fig.~\ref{fig_self_duality} we represent
the energy per vortex as a function of $\mu$. If the theory were self-dual, there would be a value
of $\mu$ where the energy per vortex number $E/n$ would not depend on $n$. We clearly see in Fig.~\ref{fig_self_duality} that value does not exist.

\subsection{Configurations with $p_{2}\neq 0$}
\begin{figure}
\includegraphics[angle=-90,width=0.5\textwidth]{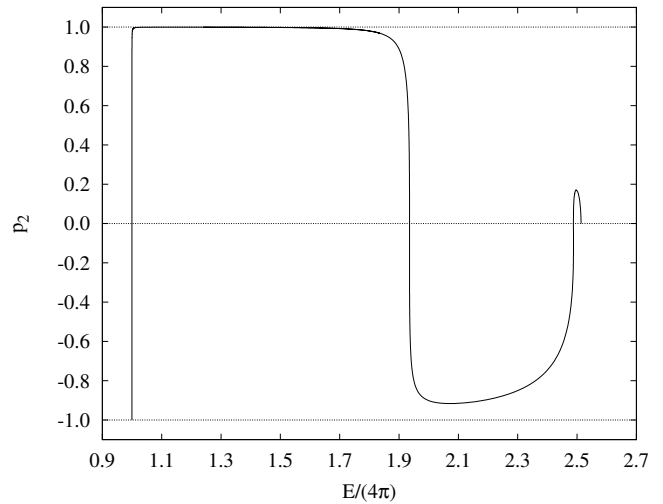}
\caption{$p_{2}$ parameter versus the energy $E$ for $n=1$, $\mu=10^{4}$ solutions.}
\label{fig_p2_E_1E4}
\end{figure}
\begin{figure}
\includegraphics[angle=-90,width=0.5\textwidth]{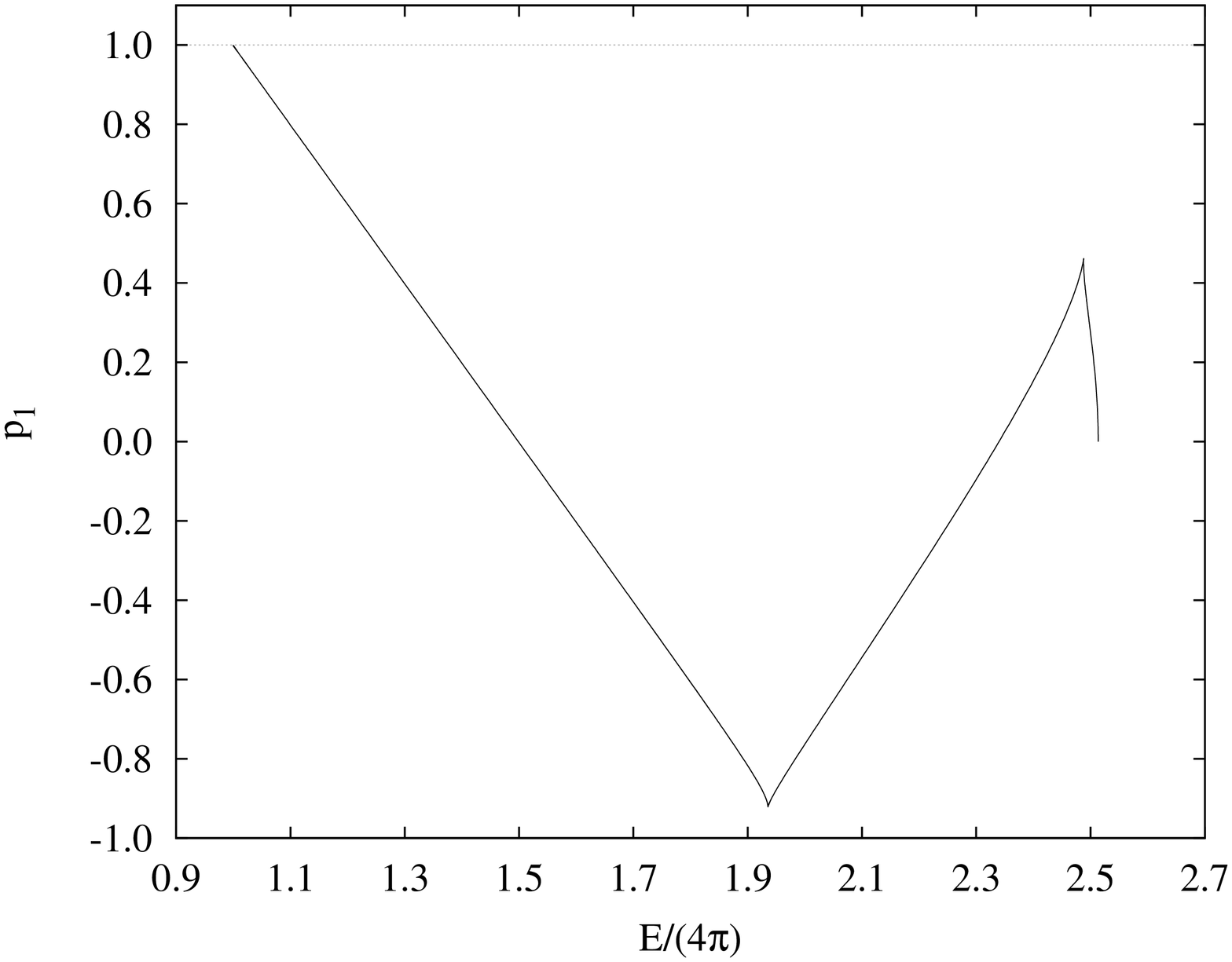}
\caption{$p_{1}$ parameter versus the energy $E$ for $n=1$, $\mu=10^{4}$ solutions.}
\label{fig_p1_E_1E4}
\end{figure}
\begin{figure}
\includegraphics[angle=-90,width=0.5\textwidth]{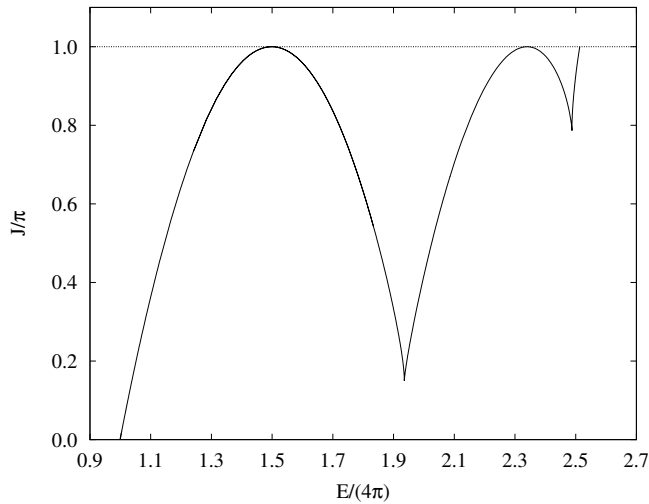}
\caption{Angular momentum $J$ versus the energy $E$ for $n=1$, $\mu=10^{4}$ solutions.}
\label{fig_J_E_1E4}
\end{figure}

In this section we explore the solutions with $p_{2}\neq 0$. These
solutions are always non-Abelian configurations and connect non-Abelian
configurations with $p_{2}=0$ to the corresponding
Abelian solutions. Note that, once a branch of configurations with parameters
$(p_{1},p_{2})$ is known, an equivalent branch can be constructed by
making an appropriate sign reverse: $(p_{1},p_{2}) \rightarrow
(-p_{1},-p_{2})$. For the sake of clarity, we will only present one of the
branches in our figures, although the other one can be obtained by mirroring the figures on the $p_{1}/p_{2}$ axis.

In Fig.~\ref{fig_p2_E_1E4} we
represent the $p_{2}$ parameter versus the energy $E$ for $n=1$, $\mu=10^{4}$ solutions. The corresponding Abelian solution possesses the highest value of $E$ (at the right end of the curve). Starting from that point we make $p_2$ deviate from zero and non-Abelian solutions ($p_{2}\neq 0$) are generated. At some point, the $p_{2}=0$ value is reached again, now corresponding to a non-Abelian solution, namely, that of the $m=3$, $p_2=0$ branch. If one continues varying $p_2$, more non-Abelian solutions are produced. Two more $p_2=0$ solutions are found on the way, corresponding to $m=2$ and $m=1$, respectively. The curve can be extended until the trivial limiting solution with $|p_2|=1$ is reached.

The other parameter we discussed, $p_{1}$, has been represented in Fig.~\ref{fig_p1_E_1E4}. It is clear that the behaviors of $p_{1}$ and
$p_{2}$ are quite different: in the regions where $p_{2}$ varies rapidly,
$p_{1}$ changes slowly, and vice versa. We take advantage of that fact from a numerical point of view, since it allows us to overcome convergence problem in certain regions of the parameter space. The relation
between the energy and $p_{1}$ gets more and more piecewise linear as $\mu$ increases.

The angular momentum $J$ is represented versus the
energy $E$ in Fig.~\ref{fig_J_E_1E4} for the same set of parameters. The
maximal value of the angular momentum $(J_{max}=\pi n^{2})$ is reached at
solutions with $p_1=0$ (see Eq.~(\ref{total_angular_momentum})). Together with
the Abelian solution, for these parameters, we have two non-Abelian solutions
with maximal angular momentum. It is interesting to realize the arch-like
structure of the figure. This is something characteristic of large values of
$\mu$ and we will discuss it in more detail later.

Although the results presented here are for $n=1$, $\mu=10^{4}$, these features are general for other values of $\mu$ and $n$. However, for $n=3$ (and beyond), we find some peculiarities we will discuss in next section.

\subsection{Uniqueness violation for $n=3$}
\begin{figure}
\includegraphics[angle=-90,width=0.5\textwidth]{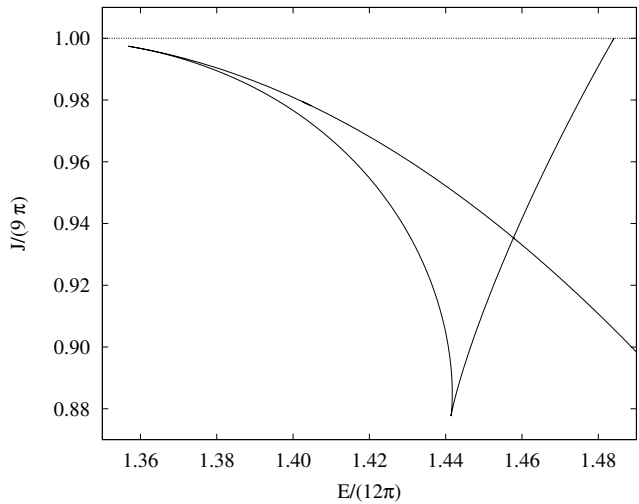}
\caption{Angular momentum $J$ versus the energy $E$ for $n=3$, $\mu=10$ (detail of the whole curve). The
  intersection of the curve shows the violation of uniqueness in configurations
  with $n=3$.}
\label{fig_unicity}
\end{figure}
\begin{figure}
\includegraphics[angle=-90,width=0.5\textwidth]{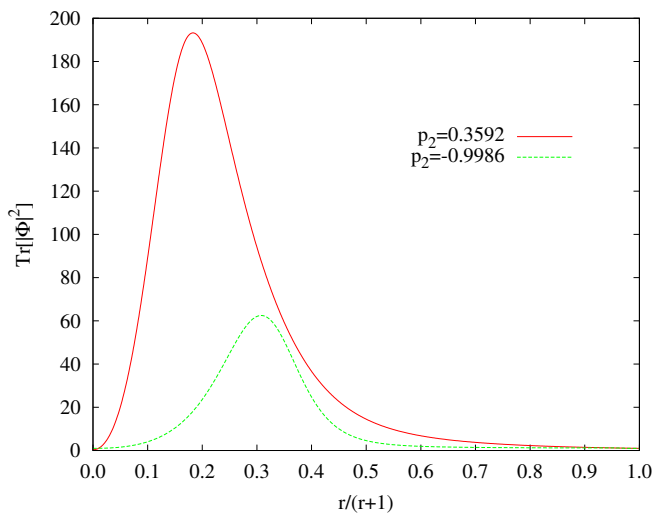}
\caption{$Tr[|\Phi|^{2}]$ versus the compactified radial coordinate for $n=3$, $\mu=10$
   for different configurations with equal energy and angular momentum
  ($E/12\pi=1.459, J/9\pi=0.9353$). 
}
\label{fig_unicity_Phi2}
\end{figure}

Even though the situation explained in last section is quite similar to
the situations found for other values of $\mu$ and $n$ (except for the number
of non-Abelian branches and the branching points) for $n=3$ a peculiar
behavior has been found for relatively low values of the coupling constant:
we find configurations with the same energy and angular momentum, but
different internal parameters.

In Fig.~\ref{fig_unicity} we present this feature for $n=3$, $\mu=10$. The
curve begins at the Abelian configuration (at the right end with maximal
angular momentum). When we move away from the Abelian solution, we obtain two non-Abelian configurations that posses the
same energy and angular momentum ($E/12\pi=1.459, J/9\pi=0.9353$), but
different internal $p_2$ parameter ($p_{2}^{(1)}=0.3592$ and
$p_{2}^{(2)}=-0.9986$). Not all the internal parameters are different: because
the angular momentum is equal for both solutions, the $p_{1}$ parameter  is also equal.

To be sure that both solutions are not related by a gauge transformation we can consider a gauge invariant quantity and evaluate it for both solutions. In Fig.~\ref{fig_unicity_Phi2} we represent the Higgs field density $Tr[|\Phi|^{2}]$ and we can see that they are clearly different solutions.

For lower values of the vorticity number, this phenomenon has not been
observed. In fact, for $n=3$ and high enough $\mu$, this phenomenon also disappears. This lack of uniqueness in certain regions of the parameter space may lead us to introduce another quantity that allows us to classify uniquely the solutions by means of $\mu$, $n$, $E$, $J$, and this new quantity. That quantity has to be gauge invariant. In order to propose one, further research needs to be done.

\subsection{The $\mu \to \infty$ limit}

\begin{figure}
\includegraphics[angle=-90,width=0.5\textwidth]{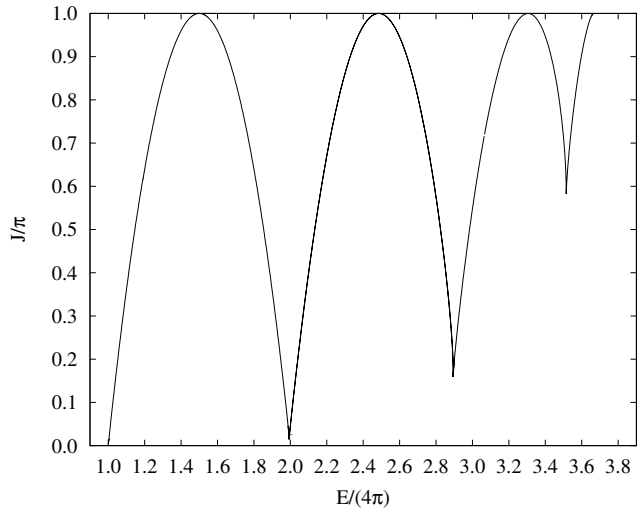}
\caption{Angular momentum $J$ versus the energy $E$ for $n=1$, $\mu=10^{6}$ solutions.}
\label{fig_J_E_1E6}
\end{figure}
\begin{figure}
\includegraphics[angle=-90,width=0.5\textwidth]{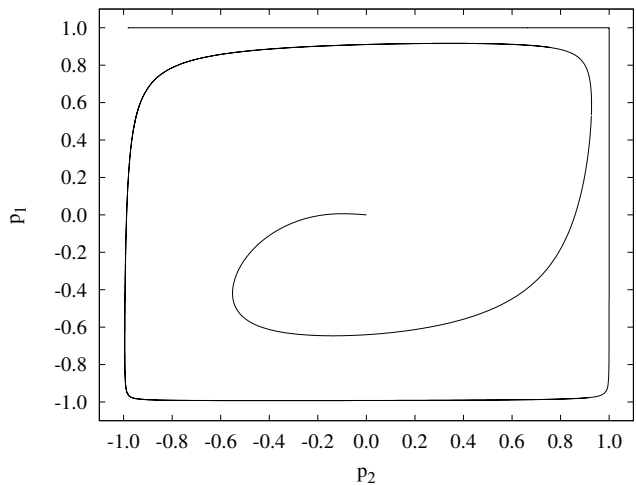}
\caption{$p_{1}$ parameter versus $p_{2}$ for $n=1$, $\mu=10^{6}$ solutions.}
\label{fig_p1_p2_1E6}
\end{figure}
\begin{figure}
\includegraphics[angle=-90,width=0.5\textwidth]{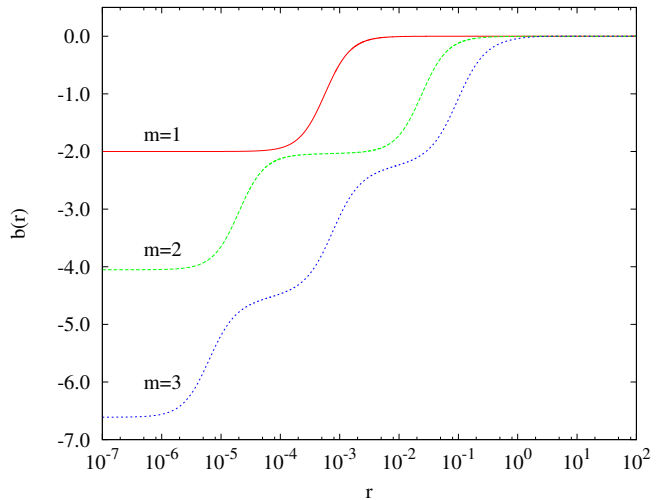}
\caption{The gauge field function $b(r)$ versus $r$ for the $n=1$, $\mu=10^{6}$,  $p_{2}= 0$, $m=1,2,3$ solutions.}
\label{fig_b_functions_1E6}
\end{figure}

In this section we will analyze the behavior of the configurations
when the Higgs self-coupling parameter tends to infinity. In that limit the theory becomes a gauged $\sigma$-model, with $Tr[F^{2}]\neq 0$.

The main feature in the $\mu \to \infty$ limit is that the energy becomes piecewise linear in $p_{1}$, which results in a piecewise quadratic relation between the
energy and the angular momentum (as a consequence of Eq.~(\ref{total_angular_momentum})). One can see this result for $n=1$,
$\mu=10^{6}$ in Fig.~\ref{fig_J_E_1E6}. The arch structure that can be seen
in this figure can be written explicitly for the general case when $\mu=\infty$:

\begin{equation}
J = n^2\pi - \frac{1}{4 \pi}\left[ E - 4 n \pi (m+\frac{1}{2})\right]^2 , \label{Regge_relation}
\end{equation}
for $E\in[4\pi m n,4\pi (m+1) n]$.


One can notice that in the limit $\mu\to\infty$ the angular momentum $J$ is a piecewise quadratic polymonial in $E$. This reminds us of Regge-like dispersion relations where the angular momentum is a function of the square of the mass (when referred to maximal angular momentum configurations.) 

In that limit, neither the energy nor the angular momentum depends on the $p_{2}$
parameter. The reason for this can be seen numerically: in the
limit $\mu \to \infty$, for $p_{2}\in(-1,1)$ we have $|p_{1}| = n$. On
the other side, for $p_{1}\in(-n,n)$, we have
$|p_{2}|=1$. So in the limit, all the solutions with $p_{2}\in(-1,1)$ are
degenerated into the minimal angular momentum solutions. We present this
result for $n=1$, $\mu=10^{6}$ in Fig.~\ref{fig_p1_p2_1E6}. Note that the
$p_{1}=p_{2}=0$ configuration corresponds to the Abelian solution. In the limit $\mu \to
\infty$ there is no such a solution (the energy of the Abelian configuration
diverges). What is left is the limiting non-Abelian configurations with
$\{|p_{1}|=n,p_{2}\in(-1,1)\} \cup \{|p_{2}|=1,p_{1}\in(-n,n)\}$.

This behavior of the configurations is closely related to the structure of the
functions when $\mu \to \infty$. In Fig.~\ref{fig_b_functions_1E6} we show
the structure for the gauge field function $b(r)$, for $n=1$, $\mu=10^{6}$, and
$m=1,2,3$, that is, the three non-Abelian $p_{2}=0$ branches.  In general, $b(r)$ presents a tendency to become overlapping step functions as $\mu \to \infty$, where the number of steps of the function is given by the branch number $m$.

\section{Summary and discussion}
In this paper we have constructed non-Abelian vortices in an $SU(2)$ CSH theory in 2+1 dimensions with a quartic Higgs potential. Contrary to what happens for HKP-JW solutions \cite{Hong,JW}, in the Abelian sector of this theory no self-dual limit is present. However, we have investigated this model since the Higgs potential we have used is the standard quartic one. This fact makes the non-Abelian solutions presented here very different from the ones described in \cite{CSH}, where a sixth-order potential was used instead.

In order to generate these solutions we start from their corresponding Abelian counterparts. We observe that for certain values of $\mu$ non-Abelian branches with $p_2=0$ appear. These non-Abelian $p_2=0$ solutions are important to understand the structure of the solution space. These solutions may be labelled by an integer number $m$, which results to be related to the steplike structure of function $b$. All these non-Abelian $p_2=0$ solutions possess lower energy than their corresponding Abelian one, something that not always happened for the sixth-order potential. The same holds for the angular momentum, which is maximal for the Abelian solutions. The structure of the energy levels of these non-Abelian $p_2=0$ solutions is quite regular, becoming completely equidistant in the limit $\mu\to\infty$.  

Apart from the Abelian and the non-Abelian $p_2=0$ solutions there are generic non-Abelian $p_2 \neq 0$ solutions which connect those both types of solutions. In fact, starting from an Abelian solution (for fixed values of $n$ and $\mu$) one can move $p_2$ and generate a whole branch of non-Abelian solutions, which ends when the limiting solution (with $|p_2|=1$) is reached.  

In this theory we observe an interesting feature for vortex number $n=3$ (and beyond): the violation of uniqueness. For $n=3$ there are regions in the parameter space where the solutions are no longer characterized by their global charges $n$, $E$, $J$, for a given $\mu$. It is possible to find different solutions with the same values for all these quantities. This lack of uniqueness brings us to look for another (gauge-invariant) quantity that allows us to characterize uniquely the solutions. Although we have some candidates we have not yet decided which one is more appropriate to do the job.

Finally, we have addressed the theory in the limit $\mu\to\infty$, when it changes from a Higgs model to an $O(3)$ gauged sigma model. In such a limit one can extract the exact relation between the energy $E$ and the angular momentum $J$, Eq.~(\ref{Regge_relation}), which results to be a Regge-like relation when the energies are referred to the energies of the solutions with maximal angular momentum ($|p_1|=n$ solutions).

\begin{acknowledgments}
We thank E. Radu for useful discussions and comments on this paper. This work was carried out in the framework of the Spanish Education and Science Ministry under Project No. FIS2011-28013 and the Science Foundation Ireland under Project No. RFP07-330PHY. J.L.B.-S. was supported by Universidad Complutense de Madrid.
\end{acknowledgments}

\bibliographystyle{physrev}

\end{document}